\documentclass[doublecol]{epl2} 

\usepackage{graphicx}
\usepackage{amsmath}
\usepackage{amssymb}
\usepackage{bm}
\usepackage{graphicx}
\usepackage{epsfig}
\usepackage{hyperref}
\usepackage[english]{babel}

\title{Competing superconducting, magnetic and charge orderings in the AF Heisenberg-Kondo lattice with Dirac electrons}
\shorttitle{Superconductivity in the Heisenberg-Kondo lattice with Dirac electrons} 

\author{E. C. Marino\inst{1} \and Lizardo H. C. M. Nunes\inst{2}}
\shortauthor{Marino \etal}

\institute{                    
  \inst{1} Instituto de F\'{\i}sica, Universidade Federal do Rio de Janeiro, Caixa Postal 68528, Rio de Janeiro, RJ, 21941-972, Brazil\\
  \inst{2} Departamento de Ci\^encias Naturais, Universidade Federal de S\~ao Jo\~ao del Rei, 36301-000 S\~ao Jo\~ao del Rei, MG, Brazil
}

\abstract{
Many recently discovered advanced materials, such as high-Tc cuprates, iron pnictides and several heavy-fermions, exhibit a rich phase diagram suggesting the presence of different competing interactions that would lead to various types of ordering. Nevertheless, there is not yet a clear unifying picture allowing the understanding of the detailed mechanisms that generate such competing interactions. Having such a picture, however, could quite well be at the very roots of the requirements for understanding high-Tc superconductivity in cuprates and pnictides, for instance. In this work we consider the antiferromagnetic (AF) Heisenberg-Kondo lattice, consisting of localized spins with AF exchange interactions between nearest neighbors on a square lattice and itinerant electrons, which undergo a magnetic Kondo interaction with the localized spins, but are otherwise non-interacting. Using the Schwinger-boson (CP$^1$) formalism and assuming the electrons are Dirac-like, we integrate on the localized degrees of freedom thereby obtaining the effective interaction among the itinerant electrons. This contains a BCS-like superconducting term, a Nambu-Jona-Lasinio-like, charge gap term and a Ising and Heisenberg-like magnetic terms. All these four competing interactions, therefore are generated by the original Kondo magnetic interaction.}

\begin{document}

\maketitle

\section{Introduction}\label{Introduction}

Many recently discovered systems in condensed matter physics
present a deep interplay among different types of orderings
such as superconducting, magnetic or charge ordering. Also in some of them
the electronic excitations behave as Dirac fermions
 These include iron pnictides, superconducting cuprates, transition metal dichalcogenides (TMD)
and possibly graphene, if the material is doped, strained or has atoms adsorbed on its surface.

Iron based pnictide materials, for instance,
undergo a transition from a magnetically ordered state to a superconducting one
upon doping~\cite{Cruz2008,Kamihara2008,Rotter2008}.
For the particular case of the 122 materials,
magnetic order and superconductivity coexist in a small region of the phase diagram~\cite{Goko2009}
and the new quasiparticles in that region
exhibit a Dirac-like linear energy dispersion~\cite{Richard2010,Khuong2011,PRL2011}
with Cooper pairing possibly promoted
by short wavelength antiferromagnetic spin fluctuations.

On the same token, the parent compounds of the cuprate superconductors are insulators presenting  AF order.
As charge carriers are added to the CuO$_{ 2 } $ planes, there is the onset of superconductivity,
with the characteristic dome-shaped superconducting phase diagram~\cite{cuprates,Nagaosa2006}.
Dirac points appear in the intersection of the nodes of the $ d $-wave superconducting gap and the two-dimensional (2D) Fermi surface~\cite{Affleck1988,Affleck1989,Wen1996}.
Strongly interacting 2D Dirac fermion systems also exhibit a dome structure in their superconducting phase diagram~\cite{Smith2009,Nunes2010}, so
we may specculate whether Dirac fermions may play any role in the description of the cuprate superconductors.

Moroever, the quasi 2D TMD are layered compounds
where $ s $-wave superconductivity coexists with a charge density wave (CDW) at low temperatures
and applied pressure~\cite{Withers1986,Wilson1969,Wilson1975}.
A theory has been proposed in which Dirac electrons appear close to the nodes of the CDW gap
and form Cooper pairs~\cite{CastroNeto2001,Uchoa2005}.
The theory is consistent with the linear decay of the temperature dependent critical field~\cite{Marino2007},
which is observed experimentally in the copper doped dichalcogenide Cu$_x$TiSe$_2$~\cite{Morosan2006}.

Graphene, on the other hand, is a semi-metal with gapless electronic quasiparticles, which
due to the peculiar lattice structure behave as Dirac fermions. Pure graphene does not exhibit
superconductivity or magnetism,
however, upon doping and/or straining may display competing orders, such as local magnetic moments, superconductivity or an excitonic gap~\cite{Kotov2011}.

Heavy fermions is another vast class of materials presenting a rich competition of different types of order in their phase diagrams~\cite{heavyfermions,Stewart2001,Stewart2006}

The above mentioned systems have been
extensively studied both theoretically and experimentally but, nonetheless, we still do not have a clear unified picture affording
a detailed understanding of the microscopic mechanisms that lead to each kind of ordering. It would be very instructive and inspiring,
therefore, to find fully controllable models for systems displaying phases with the aforementioned different types of ordering, where one could precisely trace back the original interaction and the mechanisms leading to such phases.

 In this work, we investigate the AF Heisenberg-Kondo lattice model~\cite{spin-fermion},
 a system containing both localized spins and itinerant electrons with a mutual Kondo-like magnetic interaction.
Our main aim is to determine what is the effective net interaction among the conduction electrons, which results from their magnetic interaction with the AF substrate of localized spins. For this purpose, we use the well-known
 Nonlinear Sigma Model (NLSM) description of the latter~\cite{Chakravarty89}.
The itinerant electrons, conversely, are supposed to have a tight-binding band structure
showing the presence of Dirac cones whose vertices touch at the interface between the valence and conduction bands.
We assume the system to be close to half-filling and therefore describe the kinematics of the itinerant electrons by the Dirac hamiltonian.
In order to describe the the magnetic Kondo interaction it is convenient to re-phrase the NLSM in the CP$^1$ language~\cite{Auerbach94},
whereby that interaction becomes a quartic term involving two Dirac fermion fields and two bosonic (Schwinger Boson) fields.

Our strategy will be to functionally integrate over the bosonic fields in order to derive the resulting effective interaction existing among the fermion fields. In this process, we show that the Kondo interaction among itinerant and localized electrons can be completely expressed
as a gauge coupling between the Dirac fermions and the Schwinger bosons, mediated by the CP$^1$ vector gauge field.
Our final result for the effective electron interaction is then obtained upon integration over this gauge field.

The resulting interaction possesses four pieces: a superconducting BCS-type term,
an Ising and a Heisenberg-like magnetic interactions
and a Nambu-Jona-Lasinio-type interaction.
These terms will favor respectively superconducting, magnetic and insulating charge-gapped ordering. We conclude that the original
system of localized and itinerant spins yields ultimately an interacting electronic system with those types of competing orders.

\section{The Model and Its Continuum Limit}\label{TheModel}

We consider a single layered system containing both localized and itinerant spins in which the former are located at
the sites of a square lattice and have an antiferromagnetic exchange integral while the latter are conduction electrons
with a tight-binding dispersion relation, which is assumed to be Dirac-like. The localized spins mutual interaction will be described
by an AF Heisenberg hamiltonian on a 2D square lattice, whereas their interaction with the itinerant ones, by a Kondo-like term. The complete
hamiltonian, therefore, contains three parts, namely
\begin{eqnarray}
 H
& = & J \sum_{ \langle ij \rangle } {\bf S }_i \cdot {\bf S }_j
- t \sum_{ \langle ij \rangle } \left ( c_{i\alpha}^\dagger c_{j\alpha} + hc  \right)
\nonumber \\
& &
+ J_K \sum_{i} {\bf S }_i \cdot \left ( c_{i\alpha}^\dagger {\vec \sigma}_{\alpha\beta} c_{i\beta} \right )
\, ,
\label{EqH}
\end{eqnarray}
where ${\bf S }_i$ is the localized spin operator and $c_{i\alpha}^\dagger$ is the creation operator of an itinerant electron of spin $\alpha=\uparrow,\downarrow$, both at site $ i $. Frequently we have materials for which there are electrons coming from different bands
or even from inequivalent regions of the Brillouin zone. In these cases we would add an extra label $a=1,...,N$ to the electron operators.
Here, for the sake of simplicity, we shall omit such a label, this fact having no effect in our conclusions.

In order to obtain the partition function,
we employ the continuum path integral approach. By using a basis of spin coherent states we have the localized spins $ {\bf S }_i$ replaced by
 their correspondidng eigenvalues: $S {\bf N }( {\bf x } ) $, where $S$ is the spin quantum number and
 $ | {\bf N }( {\bf x } ) |^{ 2 } = 1 $.  $ {\bf N }$ is then decomposed into two perpendicular components associated respectively
with ferromagnetic and antiferromagnetic fluctuations\cite{Tsvelik95},
\begin{equation}
{\bf N }( {\bf x } ) = a { \bf L}( {\bf x } ) + ( - 1 )^{ | {\bf x} | } \sqrt{ 1- a^{2}| {\bf L } |^{2} } \, { \bf n}( {\bf x } )
\, ,
\label{EqCoherentDecomposition}
\end{equation}
where  $ a $ is the lattice parameter. In the continuum limit ($a\rightarrow 0$), this becomes
\begin{equation}
{\bf N }( {\bf x } ) = a { \bf L}( {\bf x } ) + ( - 1 )^{ | {\bf x} | }  \, { \bf n}( {\bf x } ) + O(a^2)
\, .
\label{EqCoherentDecomposition1}
\end{equation}
Notice that we always have $ | {\bf n }( {\bf x } ) |^{ 2 } = 1 $.

In terms of these and of the continuum fermion field  $ \psi_{ \alpha }( {\bf x } ) $ corresponding to $c_{i\alpha}$
we can express the partition function as the functional integral
\begin{eqnarray}
\mathcal{Z}
& = &
\int
\mathcal{D}\psi
\,
\mathcal{D}\psi^{ \dagger}
\,
\mathcal{D}{\bf L}
\,
\mathcal{D}{\bf n} \ \delta\left[ | {\bf n } |^{ 2 } - 1 \right]
 \nonumber \\
& &
\times
\exp
\left[ - \int_0^\beta d\tau \int d^2 x \,
\left( \,
\mathcal{ H }
-
\psi^{ \dagger } i\partial_{ \tau } \psi
\, \right)
\right]
\, ,
\label{EqZ}
\end{eqnarray}
where the continuum hamiltonian density reads
\begin{eqnarray}
{\mathcal H }
& =  &
\psi^{ \dagger } \left( i \vec \sigma \cdot \vec \nabla - \mu \right)   \psi
+ \rho_{ s } | {\bf \nabla } {\bf n } |^{ 2 } + \chi_{ \perp } S^{2 } | {\bf L } |^{ 2 }
\nonumber \\
& &
+
S {\bf L } \cdot \left[ J_{ \mbox{\scriptsize{K}} }  \; {\bf s }
+
i  \, \left(  {\bf n } \times \partial_{ \tau } { \bf n}  \right)\right]
\nonumber \\
& &
+ ( - 1 )^{ | {\bf x} | } S J_{ \mbox{\scriptsize{K}}}{\bf n } \cdot {\bf s }
\, ,
\label{EqHNLSM}
\end{eqnarray}
with $ \rho_{ s } = J S^{ 2 } $ as the spin stiffness, $ \chi_{ \perp } = 4 J $ the transverse susceptibility
and $ {\bf s } $  the itinerant spin operator, given by
\begin{equation}
{\bf s }
=
 \psi^{ \dagger }_{ \delta } \left(  \vec{ \sigma } \right)_{ \delta \gamma } \psi_{ \gamma }
\, .
\label{Eqs}
\end{equation}

In  expression (\ref{EqHNLSM}), the first term is the continuum electron kinetic hamiltonian density, derived from the tight-binding
energy assuming the system has a Dirac-like dispersion relation near the Fermi points.
As the system is doped, charge carriers are added or removed from the conduction band. Their total number
is controlled by a chemical potential $\mu$, which has, therefore, been included in the previous equation.

Integrating over $ \bf L $ in (\ref{EqZ})
we obtain the resulting effective lagrangian density
\begin{eqnarray}
\mathcal{ L }_{  \mbox{\scriptsize{eff}}}
& = &
 \psi^{ \dagger } \left[ i\gamma^0\gamma^\mu\partial_\mu -\mu\right] \psi
+
\frac{ \rho_{ s }  }{ 2 }
\left(
| {\bf \nabla } {\bf n } |^{ 2 } - \frac{ 1 }{ c^{2 } } | \partial_{ t} {\bf n } |^{ 2 }
\right)
\nonumber \\
& &
\hspace{-0.7cm}
+
J_{ \mbox{\scriptsize{K}} }
\left\{( - 1 )^{ | {\bf x} | } S {\bf n } +
\frac{ i }{ \chi_{ \perp} } \left(  {\bf n } \times \partial_{ \tau } {\bf n } \right)
+  \frac{ J_{ \mbox{\scriptsize{K}} } }{ 2 \chi_{ \perp} } {\bf  s }
\right\}
\cdot {\bf  s }
\, ,
\label{EqH2}
\end{eqnarray}
 where $(\gamma^0)^2=1$, $\gamma^0\gamma^i=\sigma^i$ and
 $ c = \sqrt{ \rho_{ s } \chi_{ \perp } } $ is the spin-wave velocity.

It will be convenient to use the CP$^1$ (Schwinger Boson) formulation of the
 O(3) NLSM, in which the AF fluctuation field is written as
\begin{equation}
n_{ i } =  z^{ * }_{ \alpha } \left( \sigma_{ i } \right)_{ \alpha \beta } z_{ \beta } \, , \ \ \ \  i = x, y, z,
\,
\label{Eqz1z2}
\end{equation}
in terms of the two complex fields $ z_{ \alpha }$, $\alpha = 1, 2 $,
satisfying the constraint
 $ | z_{ 1} |^{ 2 } + | z_{ 2 } |^{ 2 } = 1 $. In the CP$^1$ language the effective lagrangian density
 (\ref{EqH2}) is rewritten as
\begin{eqnarray}
\mathcal{ L }_{  \mbox{\scriptsize{eff}}}
& = &
 \psi^{ \dagger }  \left[ i\gamma^0\gamma^\mu\partial_\mu -\mu\right] \psi
+
2 \rho_{ s } |  D_\mu z_i  |^2
\nonumber \\
& &
+
J_{ \mbox{\scriptsize{K}} }  \  \psi^{ \dagger } \left [ ( - 1 )^{ | {\bf x} | }S\  \vec{ \sigma }\cdot {\bf n }
+ \frac{ i }{ \chi_{ \perp} } \vec{ \sigma }\cdot \left( {\bf n } \times \partial_{ \tau } {\bf n } \right)\right ]
\psi
\nonumber \\
& &
+  \frac{ J_{ \mbox{\scriptsize{K}} }^2 }{ 2 \chi_{ \perp} } {\bf  s}
\cdot {\bf  s }
\, ,
\label{EqH21}
\end{eqnarray}
where the components of ${\bf n}$ are given by (\ref{Eqz1z2}) and
$D_\mu = \partial_\mu-i A_\mu$.

\section{A Gauge Coupling Replaces the Magnetic Interaction }

We now perform a canonical transformation \cite{Marino2002} on the electron field, namely
\begin{equation}
\psi_{ \alpha } \rightarrow  U_{ \alpha \beta } \, \psi_{ \beta }
\, ,
\label{EqCanonicalTransformation}
\end{equation}
where the unitary matrix $ U $ is written in terms of the $ z_{ \alpha } $-fields as
\begin{equation}
U
=
\begin{pmatrix}
  z_{ 1 }   &  - z^{ * }_{ 2 }  \\
  z_{ 2 }   &   \ \  z^{ * }_{ 1}
\end{pmatrix}
\, .
\label{EqU}
\end{equation}

This matrix has the following property
\begin{equation}
U^\dagger \vec{ \sigma }\cdot {\bf n } U = \sigma^z
\label{EqH04}
\end{equation}
and therefore the first term in the second line in (\ref{EqH21}) can be expressed, up to a sign, as the density difference of electrons
with opposite spins:
\begin{equation}
( - 1 )^{ | {\bf x} | } J_{ \mbox{\scriptsize{K}} }  S \left( \psi^\dagger_{\uparrow } \psi_{\uparrow }  -\psi^\dagger_{\downarrow } \psi_{\downarrow }
\right).
\label{1}
\end{equation}

Assuming a uniform density of electrons, we conclude that this term will vanish upon integration in  ${\bf x}$ because of the
rapidly oscillating pre-factor.

Now, since $U$ represents a local operation,
it follows that, under the transformation (\ref{EqCanonicalTransformation}), the electron kinetic term generates
the additional interaction
\begin{equation}
i \  \psi^{ \dagger }  \gamma^0\gamma^\mu \left (U^\dagger \partial_\mu U \right) \psi
\, .
\label{2}
\end{equation}

From (\ref{EqU}),
we obtain
\begin{eqnarray}
U^\dagger \partial_\mu U
=
i \sigma^z \ A_\mu
\nonumber \\
& &
\hspace{-2.7cm}
+
\begin{pmatrix}
0 &  z^*_{ 2 } \, \partial_{ \mu } z^{ * }_{ 1 } -  z^*_{ 1 } \, \partial_{ \mu } z^{ * }_{ 2 } \\
- z_{ 2 } \, \partial_{ \mu } z_{ 1 } +  z_{ 1 } \, \partial_{ \mu } z_{ 2 }   &   0
\end{pmatrix}
\, ,
\label{EqA}
\end{eqnarray}
where we used the fact that $A_\mu = -i z^{ * }_{ i } \, \partial_{ \mu } z_{ i }$ , which follows from (\ref{EqH21}).

Now, consider the polar representations of the fields $ z_{ \alpha } $,
\begin{equation}
z_{ \alpha } = \frac{ \rho_{ \alpha } }{ \sqrt{ 2 } } \, e^{ i \, \theta_{ \alpha }  }
\, , \ \ \ \  \alpha = 1, 2.
\label{EqzPolarRepresentation}
\end{equation}

Integration over the $\theta_i$ fields eliminates rapidly oscillating phase dependent terms,
such as the second term in (\ref{EqA}). We  show in the Appendix that the same happens to the
second term in the second line in (\ref{EqH21}). The effective lagrangian density, therefore, becomes
\begin{eqnarray}
\mathcal{ L }_{  \mbox{\scriptsize{eff}}}
& = &
 \psi^{ \dagger }  \left[ i\gamma^0\gamma^\mu\partial_\mu -\mu\right] \psi
+
2 \rho_{ s } |  D_\mu z_i  |^2
\nonumber \\
& &
+
 \psi^{ \dagger }_\alpha  \gamma^0\gamma^\mu  \sigma^z_{\alpha\beta} \psi_\beta A_\mu
 +  \frac{ J_{ \mbox{\scriptsize{K}} }^2 }{ 2 \chi_{ \perp} } {\bf  s} \cdot {\bf  s }
\, .
\label{EqH22}
\end{eqnarray}

Observe that the magnetic interaction between the itinerant electrons and the localized spins
manifests ultimately as a gauge coupling between the electrons and the Schwinger boson fields,
mediated by the CP$^{1} $  vector field $A_\mu$, which becomes a gauge field.
Indeed, (\ref{EqH22}) is invariant under the gauge transformation
\begin{eqnarray}
\psi & \rightarrow & e^{i \Lambda} \psi
\, ,
\nonumber \\
\theta_i & \rightarrow & \theta_i + \Lambda
\, ,
\nonumber \\
A_\mu & \rightarrow & A_\mu - \partial_\mu \Lambda
\, .
\label{3}
\end{eqnarray}

Our aim is to obtain the net effective interaction among the conduction electrons, associated to the fermion fields
$\psi$. For this purpose we are going to functionally integrate over the  CP$^{1} $ fields, in order to derive
the resulting effective interaction. Before doing that, however we shall
express the effective lagrangian in an explicitly gauge invariant way.

We first introduce the gauge invariant phase-fields \cite{ijmp}
\begin{equation}
\chi_i = \theta_i + \frac{\partial_\mu A^\mu}{\Box} ,
\label{4}
\end{equation}
which are clearly invariant under (\ref{3}). In (\ref{4}),
 $1/\Box$ is the Green function of the $\Box =  \partial_\mu\partial^\mu$ operator.

We can now re-write the effective lagrangian density (\ref{EqH22})
in an explicitly gauge invariant form given by \cite{ijmp}
\begin{eqnarray}
\mathcal{ L }_{  \mbox{\scriptsize{eff}}}
& = &
 \psi^{ \dagger }  \left[ i\gamma^0\gamma^\mu\partial_\mu -\mu\right] \psi
+
\frac{1}{2} \sum_{i=1}^2\ \rho_i^2\ \partial_\mu \chi_i \partial^\mu \chi_i
\nonumber \\
& &
+ \frac{1}{4} F_{\mu\nu}\left [\frac{4 \rho_s}{-\Box} \right ] F^{\mu\nu}
+ \psi^{ \dagger }_\alpha  \gamma^0\gamma^\mu  \sigma^z_{\alpha\beta} \psi_\beta A_\mu
\nonumber \\
& &
+ \frac{ J_{ \mbox{\scriptsize{K}} }^2 }{ 2 \chi_{ \perp} } {\bf  s}
\cdot {\bf  s }
\, .
\label{5}
\end{eqnarray}
where we have used the constant $\rho_i$ approximation for $i=1,2$
and  $F_{\mu\nu}= \partial_\mu A_\nu - \partial_\nu A_\mu $.

\section{The Effective Electron Interaction}

Let us now perform the functional integration over the bosonic fields
$\rho$, $\chi$ and $A_\mu$. This will ultimately generate the final
effective interaction among the conduction electrons.

We shall adopt the constant $\rho_i$  ($i=1,2$)
approximation for performing the functional integration over the Schwinger boson fields, namely,
$\rho_i$'s and $\chi_i$'s. This approximation usually reproduces the physical situation found in many materials. It implies that integration over the $z_i$ fields would just yield a trivial multiplicative constant in the partition function.
The nontrivial interaction effect comes from integration over the gauge field $A_\mu$. This can be easily performed
given the quadratic dependence of (\ref{5}) in this field.

The resulting effective lagrangian density for the conduction electrons is
\begin{eqnarray}
\mathcal{ L }_{  \mbox{\scriptsize{eff}},\psi}
& = &
\psi^{ \dagger }  \left[ i\gamma^0\gamma^\mu\partial_\mu -\mu\right] \psi
\nonumber \\
& &
+
\frac{1}{8 \rho_s}
\left( \psi^{ \dagger }_\alpha  \gamma^0\gamma^\mu  \sigma^z_{\alpha\beta} \psi_\beta \right)
\left( \psi^{ \dagger }_\alpha  \gamma^0\gamma_\mu  \sigma^z_{\alpha\beta} \psi_\beta \right)
\nonumber \\
& &
+  \frac{ J_{ \mbox{\scriptsize{K}} }^2 }{ 2 \chi_{ \perp} } {\bf  s} \cdot {\bf  s }\
\, .
\label{6}
\end{eqnarray}

Explicitly writing the components of the Dirac field we may, after some algebra, express the effective
interaction term above as
\begin{eqnarray}
\mathcal{ L }_{  \mbox{\scriptsize{I}},\psi}
& =  &
\frac{1}{4 \rho_s}
\left(
\psi^\dag_{1\uparrow } \ \psi^\dag_{2\downarrow } + \psi^\dag_{2\uparrow } \ \psi^\dag_{1\downarrow }
\right)
\left(
\psi_{2\downarrow } \ \psi_{1\uparrow }
+ \psi_{1\downarrow } \ \psi_{2\uparrow }
\right )
\nonumber \\
& &
+ \frac{1}{8 \rho_s} s_z^2
+\frac{1}{8 \rho_s}\left [ \left (\bar \psi \psi \right )^2 - \left (\bar \psi \gamma^0 \psi \right )^2\right ]
\nonumber \\
& &
+  \frac{ J_{ \mbox{\scriptsize{K}} }^2 }{ 2 \chi_{ \perp} } {\bf  s} \cdot {\bf  s }
\, .
\label{7}
\end{eqnarray}

The first term above is a superconducting, $ s $-wave BCS-type interaction. It would lead to a superconducting phase
with an order parameter
$\Delta = \langle \psi_{1\uparrow}^\dagger \psi_{2\downarrow}^\dagger + \psi_{2\uparrow}^\dagger \psi_{1\downarrow}^\dagger\rangle$.
The second term is a Nambu-Jona-Lasinio-type interaction, that may produce an insulating charge-gapped phase showing
an excitonic condensate. The gap parameter would be
$  M = \langle \bar\psi \psi\rangle $.
Finally, the second and fourth terms in (\ref{7}) are, respectively, Ising and Heisenberg-like, magnetic interactions. These
may potentially lead to a magnetically ordered phase with the nonzero order parameter being the average magnetization
$  \vec m = \langle \vec s\rangle $.


A model containing the interactions of the first and third terms in (\ref{7}) has been investigated in a mean-field approximation~\cite{Nunes2010}.
We have shown that,
even if the excitonic interaction strength is larger than the superconducting interaction,
as the chemical potential increases, superconductivity eventually suppresses the excitonic order parameter,
which means that the system goes from an insulating state to a superconducting one
as charge carrriers are added to the system.

A model with Dirac fermions subject to the interaction described by the first term in (\ref{7}) was studied in~\cite{Marino2006,Marino2007}.
\section{Conclusions}\label{Conclusions}

We have provided a concrete example of a model containing both localized spins and itinerant electrons, where different competing interactions are generated out of the original purely magnetic interactions.
The localized spins
present an AF Heisenberg interaction on a square lattice, the itinerant electrons are Dirac-like and the mutual localized-itinerant interaction is magnetic, Kondo-like.

Our results yield an unified, controllable picture for the the common magnetic origin of three competing types of order in a strongly correlated system: superconducting, magnetic and charge ordering.

\section{Appendix}\label{Appendix}

Let us show here that the second term in the second line in (\ref{EqH21}) only contains rapidly oscillating phase dependent terms,
which are, therefore, eliminated through functional integration over the phase fields.

We may write this term as
\begin{eqnarray}
\left (  \psi^\dagger \vec{ \sigma } \psi  \right) \cdot \left( {\bf n } \times \partial_{ \tau } {\bf n } \right )
& = &
\psi^\dagger_\alpha  \psi_\beta  z^*_\mu z_\nu \partial_\tau \left (z^*_\lambda z_\rho \right )
\nonumber \\
& &
\times  \epsilon^{ijk} \sigma^i_{\alpha\beta} \sigma^j_{\mu\nu}  \sigma^k_{\lambda\rho}
\, ,
\label{7a}
\end{eqnarray}
where we have used (\ref{Eqz1z2}).

Now, we have the following identity for Pauli matrices
\begin{eqnarray}
\sigma^i_{\alpha\beta} \sigma^j_{\mu\nu}
& = &
\frac{\delta^{ij}}{3}
\left [  2 \delta_{\alpha\nu} \delta_{\beta\mu}- \delta_{\alpha\beta} \delta_{\mu\nu} \right]
\nonumber \\
& &
+ i \epsilon^{ijk} \left [ \delta_{\beta\mu} \sigma^k_{\alpha\nu}
- \delta_{\alpha\mu} \sigma^k_{\beta\nu}\right ]
\, .
\label{8}
\end{eqnarray}

Inserting (\ref{8}) in (\ref{7a}), we immediately obtain
\begin{equation}
 {\bf s} \cdot \left( {\bf n } \times \partial_{ \tau } {\bf n } \right )
  =
  \psi^\dagger_\alpha  \psi_\alpha  z^*_\mu z_\mu \partial_\tau \left (z^*_\lambda z_\lambda \right )
+ pdt ,
\label{9}
\end{equation}
where $pdt$ stands for ``phase dependent terms''.

The first term on the r.h.s. above vanishes because $ z^*_\lambda z_\lambda = 1$ and, therefore, only phase
dependent terms are left, as we have asserted.

\acknowledgments
This work has been supported by CNPq, FAPERJ and FAPEMIG.
We would like to thank H. Caldas, A. L. Mota, R. L. S. Farias and M. B. Silva Neto for discussions.

\end{document}